\documentclass[epsf,seceq,preprint]{jpsj2}

\usepackage{amssymb}

\usepackage{amsmath}

\title{Elastic properties and magnetic phase diagrams of \\
  dense Kondo compound Ce$_{0.75}$La$_{0.25}$B$_6$}

\author
{
Osamu {\sc SUZUKI},
Shintaro {\sc NAKAMURA}$^1$,
Mitsuhiro {\sc AKATSU}$^2$\thanks{Present address: National Institute for Advanced Industrial Science and Technology, Tsukuba.},\\ Yuichi {\sc NEMOTO}$^2$, Terutaka {\sc GOTO}$^2$
and Satoru {\sc KUNII}$^3$
}

\inst
{
National Institute for Materials 
Science, Tsukuba 305-0047\\
$^1$Center for Low Temperature Science, Tohoku University, Sendai 980-8577\\
$^2$Graduate School of Science and Technology, Niigata University, Niigata 
950-2181\\
$^3$Department of Physics, Tohoku University, Sendai 980-8578
}

\recdate{\today}

\abst
{
We have investigated the elastic properties of the cubic dense Kondo compound 
Ce$_{0.75}$La$_{0.25}$B$_6$ by means of ultrasonic measurements. We have 
obtained magnetic fields vs. temperatures ($H-T$) phase diagrams under 
magnetic fields along the crystallographic [001], [110] and [111] axes. An 
ordered phase IV showing the elastic softening of  $c_{44}$ locates in low 
temperature region between 1.6 K and 1.1 K below 0.7 T in all field 
directions. The phase IV shows an isotropic nature  with regard to the field 
directions, while  the antiferro-magnetic phase III shows an anisotropic 
character. A remarkable softening of $c_{44}$ and a spontaneous trigonal 
distortion $\varepsilon_{yz}+\varepsilon_{zx}+\varepsilon_{xy}$ recently 
reported by Akatsu {\it et al}. in the phase IV  favor a ferro-quadrupole 
(FQ) moment of $O_{yz}$+$O_{zx}$+$O_{xy}$ induced by an octupole ordering.
}

\kword{Ce$_{0.75}$La$_{0.25}$B$_6$, elastic constant, ultrasonic 
measurement, quadrupolar ordering, octupolar ordering, magnetic phase diagram, dense Kondo 
effect}

\begin{document}

\sloppy

\maketitle

\section {Introduction}

  CeB$_{6}$ and its substitutional compound Ce$_{x}$La$_{1-x}$B$_6$ are well 
known as the typical examples showing multipolar orderings and the dense 
Kondo effect. In CeB$_{6}$ the Hund's ground multiplet $^2F_{5/2}$ of 
Ce$^{3+}$ splits into a quartet $\Gamma_{8}$ ground state and a doublet 
$\Gamma_{7}$ excited state located at 540 K in the cubic 
crystal-electric-field 
(CEF)~\cite{rf:Zirngiebl1984PRB,rf:Luthi1984ZPhys,rf:Loewenhaupt1985JMMM,rf: 
Sato1984JPSJ}. The CEF ground state of Ce$_{x}$La$_{1-x}$B$_{6}$ is also 
considered to be a quartet $\Gamma_{8}$  from the results of the magnetic susceptibility measurements~\cite{rf:Sato1984JPSJ}. Because of 
the large CEF splitting energy, the low-temperature properties of 
Ce$_{x}$La$_{1-x}$B$_6$ are dominated by the ground state quartet 
$\Gamma_{8}$ with SU(4) symmetry. The direct product $\Gamma_{8}\otimes\Gamma_{8}$ possesses 
fifteen quantum degrees of freedom consisting of  three magnetic dipoles, 
five electric quadrupoles, and seven magnetic 
octupoles~\cite{rf:Shiina1997JPSJ}. Actually CeB$_6$ undergoes successive 
transitions from higher temperature paramagnetic phase I into an 
antiferro-quadrupolar (AFQ) phase II at $T_{\rm Q}=3.3$ K and further into 
an antiferro-magnetic (AFM) phase III below $T_{\rm N}=2.3$ K in which the 
AFQ ordering coexists~\cite{rf:Fujita1980SSC,rf:Effantin1985JMMM}.

  For a few decades, the AFQ phase II of CeB$_6$ has attracted much 
attention because of its order parameter and unusual magnetic field dependence of the AFQ 
transition temperature $T_{\rm Q}$, whose behavior cannot be interpreted 
in terms of the conventional AFM ordering. With increasing magnetic fields, 
the transition point $T_{\rm Q}$ for the AFQ phase II shifts to higher 
temperatures. The result of neutron diffraction on CeB$_6$ indicates a 
field-induced AFM moments with a single wave vector 
$\mib{k}=\left[\frac{1}{2},\frac{1}{2},\frac{1}{2}\right]$ in the phase II, 
which was consider to be evidence of the AFQ 
ordering~\cite{rf:Effantin1985JMMM}. However, the internal magnetic field 
in the phase II obtained by $^{11}$B NMR measurements was interpreted as a 
triple-${\mib k}$ AFM structure.~\cite{rf:Takigawa1983JPSJ}
In order to solve this inconsistency between the neutron diffraction and 
NMR experiments, the influence of a field-induced antiferro-octupole (AFO) 
has been taken into consideration.~\cite{rf:Sakai1997JPSJ}
The splitting and angular dependence of NMR signals have been explained by 
the hyperfine coupling between the nuclear spin of $^{11}$B and the 
magnetic octupole moment of $T_{xyz}$ of Ce 4$f$-electron possessing the 
anti-parallel-arrangement as same as the $O_{xy}$-AFQ ordering with a wave 
vector $\mib{k}=\left[\frac{1}{2},\frac{1}{2},\frac{1}{2}\right]$. The 
mean-field calculation including the AFO interaction by Shiina {\it et al.} 
reproduces well the experimental observation that the AFQ transition point 
$T_{\rm Q}$ increases considerably in magnetic fields. Their calculation 
indicates that the AFQ phase II is indeed stabilized by the AFO interaction 
in high magnetic fields.~\cite{rf:Shiina1997JPSJ,rf:Shiina1998JPSJ,rf:HiroiPRL,rf:Hall,rf:AkatsuPRL}

  The substitution of non-magnetic lanthanum ions for magnetic cerium ones 
in Ce$_{x}$La$_{1-x}$B$_6$ modifies drastically the magnetic phase diagram 
at low temperatures (low-$T$) in low magnetic field (low-$H$) region. The 
quadrupolar interaction for the AFQ phase II reduces much faster than the 
magnetic dipolar interaction for the AFM phase III as reducing the cerium 
concentration $x$. A new ordered phase IV manifests itself in the compound 
Ce$_{0.75}$La$_{0.25}$B$_6$ in addition to the AFQ phase II and AFM phase 
III.~\cite{rf:Suzuki1998JPSJ,rf:Tayama1997JPSJ,rf:Hiroi1997PRB}
The specific heat of Ce$_{0.75}$La$_{0.25}$B$_6$ in the zero magnetic field 
shows a sharp peak at $T_{\rm c1}$ $\sim$ 1.6 K and a small one at $T_{\rm 
c2}$ $\sim$ 1.1 K.
The magnetic susceptibility in the phase IV of Ce$_{0.75}$La$_{0.25}$B$_6$ 
is quite isotropic with respect to the magnetic field directions along the 
three principal [001], [110] and [111] axes~\cite{rf:Tayama1997JPSJ}. It is 
difficult to explain these unusual characteristics in terms of the 
conventional AFM ordering.

Recently, several microscopic measurements have 
been performed on Ce$_{x}$La$_{1-x}$B$_6$ in the phase IV. No indication of 
AFM ordering in the phase IV was found by the neutron scattering of powder 
samples~\cite{rf:Iwasa2003PhysicaB}. The NMR and muon spin relaxation measurements on Ce$_{x}$La$_{1-x}$B$_{6}$ ($x$ $\sim$ 0.7) indicate the breaking of the time reversal symmetry in the 
phase IV.~\cite{rf:Takagiwa,rf:Magishi}
  Quite recently, Akatsu {\it et} {\it al.} measured the thermal expansion 
of Ce$_{x}$La$_{1-x}$B$_6$ ($x$ = 0.75, 0.70) at low 
temperatures.~\cite{rf:Akatsu2002} They successfully found that in the 
phase IV the lattice expands along the [001] axis, 
while shrinks along the [111] axis. This 
result provides an evidence for a trigonal lattice distortion due to the spontaneous strain 
$\left<\varepsilon_{yz}\right>$ = $\left<\varepsilon_{zx}\right>$ = 
$\left<\varepsilon_{xy}\right>$ = $5\times10^{-6}$ in the phase IV below $T_{c1}$. 
This tinny but definite distortion promises the fact that the 
ferro-quadrupole moment $\left<O_{yz}\right>$ = $\left<O_{zx}\right>$ = 
$\left<O_{xy}\right>$ is relevant in the phase IV. The phase IV of 
Ce$_{0.70}$La$_{0.30}$B$_{6}$ below $T_{\rm c1}$ $\sim$ 1.4 K is stable 
down to absolute zero without showing the AFM phase 
III.~\cite{rf:Tayama1997JPSJ,rf:Nemoto2002PhysicaB}

    In the more diluted system Ce$_{x}$La$_{1-x}$B$_{6}$ at $x\sim0.60$, the 
inter-site interactions for the AFQ phase II and the AFM phase III 
considerably reduces, although the Kondo temperature ($T_{\rm K}$) of about 
1 K is almost independent of the cerium concentration $x$~\cite{rf:Sato1985JPSJ}. In the 
compounds $x$ = 0.60 and 0.50, the dense Kondo effect, therefore, may play 
a dominant role to realize the non-magnetic state in low$-T$, low$-H$ region. 
~\cite{rf:Nakamura2000PRB}

The anisotropy of the elastic properties of Ce$_{0.75}$La$_{0.25}$B$_{6}$ 
in magnetic fields has not been investigated in detail so far.
  In order to elucidate the characteristics of the phase IV, we have 
investigated the elastic properties and magnetic phase diagram of 
Ce$_{0.75}$La$_{0.25}$B$_{6}$ in magnetic fields along the three principal 
[001], [111] and [110] axes by ultrasound measurements.  We also discuss 
the order parameter in the phase IV from the elastic properties and magnetic 
phase diagrams of Ce$_{0.75}$La$_{0.25}$B$_6$.
 
\section{Experimental details}

  A single crystal of Ce$_{0.75}$La$_{0.25}$B$_6$ was grown by the floating 
zone method. We cut the specimen of Ce$_{0.75}$La$_{0.25}$B$_6$ into a 
$4.8\times5.3\times6.0$ mm$^3$ rectangular shape. The specimen used in the 
present experiment is the same one used in ref. 10. We measured  elastic 
constants by means of an ultrasonic apparatus based on the phase comparison 
technique~\cite{rf:Luthi1980}. The parallel faces of the specimen 
perpendicular to the high symmetry [001], [111] and [110] axes were 
polished by fine powder of carborundum. A pair of ultrasonic transducers 
made of LiNbO$_{3}$ plates were bonded on the parallel end faces of the 
specimen by RTV silicone rubber (Shin-Etsu Chemical Co., Ltd.). The 
propagation vector $\mib k$ and the polarization one $\mib u$ of sound wave 
are chosen as $\mib k$//$\mib u$//[001] for $c_{11}$, $\mib k$//[110] and 
$\mib u$//[1${\bar 1}$0] for $(c_{11}-c_{12})/2$, $\mib k$//[001] and $\mib 
u$//[100] for $c_{44}$, respectively. We used a longitudinal ultrasonic wave with 8 MHz and a transverse one with 10 MHz. We employed a sample-immersed 
$^3$He-evaporation refrigerator equipped with a 12 T-superconducting magnet 
(Oxford Instruments Co., Ltd.) for the low-temperature measurements in 
magnetic fields.

\section{Experimental Results}

The temperature dependences of $c_{44}$ in Ce$_{0.75}$La$_{0.25}$B$_6$ 
under various magnetic fields along [001], [110], and [111] axes are shown 
in Figs. 1-3, respectively. In Fig. 1, we cited the previous results in 
fields along the [001] axis for comparison~\cite{rf:Suzuki1998JPSJ}.
In the paramagnetic phase I under zero magnetic field, $c_{44}$ softens 
slightly due to the quadrupole-strain interaction with the antiferro-type 
quadrupole inter-site interaction $g'_{\Gamma_{5}} = -2$ K~\cite{rf:Suzuki1998JPSJ}. $c_{44}$ 
turns to huge softening in phase IV below $T_{\rm c1}$ ($\sim$ 1.6 K) and 
increases rapidly below $T_{\rm c2}$ ($\sim$ 1.1 K). Here $T_{\rm c1}$ and 
$T_{\rm c2}$ denotes the transition temperature from the paramagnetic phase I 
to the phase IV and that from the phase IV to the AFM phase III", 
respectively. We identify the transition point $T_{\rm c1}$ with the onset of 
the softening of $c_{44}$ and $T_{\rm c2}$ with the increase of $c_{44}$. 
With increasing magnetic fields, the softening of $c_{44}$ reduces rapidly 
and $T_{\rm c2}$ shifts to higher temperatures. At $H$= 1.0 T, 
the softening in $c_{44}$ almost vanishes. This means that the phase IV 
disappears and another phase II emerges. A small dip is observed at the AFM 
transition point $T_{\rm N}$ in three magnetic field directions as shown in 
Figs. 1-3.

The transverse $(c_{11}-c_{12})/2$ mode also exhibits anomalies around the
I-IV phase transition point.
Fig. 4 shows the temperature dependence of $(c_{11}-c_{12})/2$ of 
Ce$_{0.75}$La$_{0.25}$B$_6$ under various magnetic fields along the [110] 
axis. In zero magnetic field, $(c_{11}-c_{12})/2$ shows a slight softening 
in paramagnetic phase I toward a sharp minimum at $T_{\rm c1}$ $\sim$ 1.6 
K. $(c_{11}-c_{12})/2$ also shows a bending at $T_{\rm c2}$ in zero 
magnetic field, which corresponds to the transition from the phase IV to 
the AFM phase III". No softening of $(c_{11}-c_{12})/2$ in the phase IV 
 is in contrast to the huge softening of  $c_{44}$. With 
increasing magnetic fields up to 0.7 T, the temperature range of the phase 
IV becomes narrower. Under the fields of 1.0 T and 1.5 T, the phase IV 
disappears. $(c_{11}-c_{12})/2$ shows a small bending at the AFQ 
transition points $T_{\rm Q}$. A sharp reduction of $(c_{11}-c_{12})/2$ at the transition point to 
the AFM phase III' and an increase of $(c_{11}-c_{12})/2$ at the transition point to the phase III 
have clearly been observed.

The temperature dependence of the longitudinal $c_{11}$ of 
Ce$_{0.75}$La$_{0.25}$B$_{6}$ under magnetic fields along the [111] axis is 
shown in Fig. 5.
The transition temperature  $T_{\rm c2}$ indicated by downward arrows shifts 
to higher temperatures with increasing magnetic fields. In contrast to 
$T_{\rm c2}$, $T_{\rm c1}$ indicated by upward arrows is almost independent 
of magnetic field.  Under the field of 2.0 T, the phase IV disappears and 
the AFQ phase II appears. Similar to the case of $(c_{11}-c_{12})/2$, 
$c_{11}$ shows a small decrease in the phase IV.

Fig. 6 shows the field dependence of $(c_{11}-c_{12})/2$ of 
Ce$_{0.75}$La$_{0.25}$B$_6$ along the  [110] axis at various fixed
temperatures. At temperatures higher than 1.6 K, the inflection in 
$(c_{11}-c_{12})/2$ indicated by arrows shows the transition points $T_{\rm 
Q}$ from the paramagnetic phase I to the AFQ phase II. In the temperature 
range 1.1 K $<$ $T$ $<$ 1.6 K, Ce$_{0.75}$La$_{0.25}$B$_{6}$ undergoes a 
successive transition from the phase IV to the AFM phase III, AFM sub-phase 
III' and AFQ phase II with increasing magnetic fields. The observed 
hysteresis in the field dependence at 1.4 K indicates the transition between 
the phase IV and the phase III to be of first-order as reported by magnetic 
susceptibility measurements by Tayama {\it et.} {\it 
al.}~\cite{rf:Tayama1997JPSJ} At 1.21 K, the phase III" was found in a 
very narrow field range. The results of $c_{44}$ in Fig. 1 and $c_{11}$ in 
Fig. 5 have already shown that the phase III" appears in zero magnetic 
field below 1.1 K. At the lowest temperature 0.54 K, 
Ce$_{0.75}$La$_{0.25}$B$_6$ exhibits the successive transitions from the 
AFM sub-phase III" to the AFM phase III indicated by a hysteretic profile 
below 1 T. A step-like increase at around 2 T corresponds to the AFM 
sub-phase III'. At around 2.5 T, $(c_{11}-c_{12})/2$ shows a 
discontinuous increase, which indicates the transition from the AFM 
phase III' to AFQ phase II.

In Fig. 7, we show the temperature dependence of $c_{44}$ in 
Ce$_{0.75}$La$_{0.25}$B$_6$ in fields above 6 T applied parallel to the [110] axis. 
Sharp bending points in $c_{44}$ are the indication of $T_{\rm Q}$ for the 
AFQ transition from the paramagnetic phase I. With decreasing temperature, 
$c_{44}$ shows the Curie-Weiss-type softening with a negative (antiferro) inter-site 
quadrupole interaction~\cite{rf:Suzuki1998JPSJ}. $c_{44}$ shows upturns 
at $T_{\rm Q}$ indicated by arrows. The increase of $c_{44}$ below $T_{\rm 
Q}$ in Fig. 7 is a common feature associated with the AFQ ordering. The AFQ 
transition temperature $T_{\rm Q}$ shifts to higher temperatures with 
increasing magnetic fields up to 12 T.

Next we show the temperature dependence of $c_{11}$ of 
Ce$_{x}$La$_{1-x}$B$_{6}$ in fields above 5 T along the [111] axis in Fig. 8. The 
sharp bending points of $c_{11}$ indicated by arrows in Fig. 8 correspond to the transition points $T_{\rm Q}$ from the phase I to the AFQ phase II. The 
transition temperature $T_{\rm Q}$ shifts to higher temperatures with 
increasing field.

Taking together the above-mentioned results, we obtained the magnetic 
phase diagrams of Ce$_{0.75}$La$_{0.25}$B$_6$ under the magnetic fields 
parallel to the three principal axes as shown in Fig. 9. In these phase 
diagrams, the AFQ phase II spreads to high fields, the AFM phase III, III' 
and III" occupies the low-$T$, low-$H$ region, while the phase IV appears 
only in narrow temperature regions under low magnetic fields. The boundary 
$T_{\rm Q}(H)$  between the phase I and the phase II for $\mib{H}$//[001] 
stays slightly lower temperature regions than those under $\mib{H}$//[111] 
and $\mib{H}$//[110]. In fields of $\mib{H}$//[111] and $\mib{H}$//[110]   $T_{\rm Q}(H)$ are quite close to each other. This anisotropy of the 
AFQ phase II is consistent with the theoretical calculation on CeB$_{6}$ in 
low-$H$ region.~\cite{rf:Shiina1997JPSJ,rf:Shiina1998JPSJ} The critical 
field between the AFM phase III' and the AFQ phase II at the lowest 
temperature in the present experiments is in order of $H_{[001]}^{\rm 
III'-II} > H_{[111]}^{\rm III'-II} > H_{[110]}^{\rm III'-II}$. Regardless 
of the field direction, the AFM phase below $T_{\rm c2}$ consists of the 
three sub phases III", III and III' from the lower field side.

\section{Discussion}

In the present study, we found the enormous softenings of $c_{44}$ in the 
phase IV of Ce$_{0.75}$La$_{0.25}$B$_{6}$ under magnetic fields along the 
three crystallographic [001], [111] and [110] axes as shown in Figs. 1-3. 
On the other hands, other elastic modes $(c_{11}-c_{12})/2$ in Fig. 4 and 
$c_{11}$ in Fig. 5 show a hardening or temperature independent behavior in 
the phase IV . The behavior of $c_{44}$ around $T_{\rm c1}$ is 
distinguished from those around $T_{\rm Q}$ to the AFQ phase II (a 
small dip) or around $T_{c2}$ to the AFM+AFQ phase III (a rapid increase). At 
the transition point from the paramagnetic phase I to the AFQ phase II, the 
elastic constants $c_{11}$ and $(c_{11}-c_{12})/2$ show a sharp bending 
(Figs. 5, 7 and 8). However, $c_{44}$ of 
Ce$_{0.75}$La$_{0.25}$B$_6$ in Figs. 1-3 shows the rapid decrease below 
$T_{\rm c1}$ with lowering temperatures in the absence of magnetic fields.
$c_{44}$ is described by the quadrupolar susceptibility associated with the 
quadrupole moments $O_{yz}$, $O_{zx}$ and $O_{xy}$. The remarkable softening of 
$c_{44}$  in the phase IV of Ce$_{0.75}$La$_{0.25}$B$_{6}$ is strictly 
distinguished from the bending of $c_{44}$  at the transition point to the 
AFQ phase II in CeB$_{6}$.

Recent thermal expansion measurement by Akatsu et al. have shown that the 
trigonal distortion $\left<\varepsilon_{yz}\right>$ = 
$\left<\varepsilon_{zx}\right>$ = $\left<\varepsilon_{xy}\right>$ = $5\times10^{-6}$ 
manifests itself in the phase IV of Ce$_{x}$La$_{1-x}$B$_{6}$ with $x$ = 0.75 
and 0.70 ~\cite{rf:Akatsu2002}. This result means that the 
ferro-quadrupole moment $\left<O_{yz}\right>$ = $\left<O_{zx}\right>$ = 
$\left<O_{xy}\right>$ with $\Gamma_{5}$  symmetry is relevant in the phase 
IV. The FQ ordering, namely the cooperative Jahn-Teller transition, is one 
of the plausible models to bring about the trigonal lattice distortion in general. In 
the case of the FQ ordering, however, a considerable elastic softening 
above $T_{\rm Q}$ is expected as a precursor of the elastic instability. As 
shown in Figs. 1-3, $c_{44}$  of Ce$_{0.75}$La$_{0.25}$B$_{6}$ shows 
the softening of only 1 $\%$ above the transition point $T_{\rm c1}$ to the 
phase IV. Furthermore, the magnetic susceptibility is expected to be silent 
at the FQ transition point. The conventional FQ 
transition is unfavorable for the cusp of the magnetization upon the 
transition to the phase IV in the experimental results by Tayama et 
al.~\cite{rf:Tayama1997JPSJ} .

Very recently Kubo and Kuramoto proposed an octupole-ordering model to 
explain the unusual elastic and magnetic properties of the phase IV in 
Ce$_{x}$La$_{1-x}$B$_{6}$~\cite{rf:KuboKura,rf:KuboKura2}. This model shows that the antiferro-octupole-ordering of $T_{x}^{\beta}+T_{y}^{\beta}+T_{z}^{\beta}$
with $\Gamma_{5u}$ symmetry in their notation simultaneously induces the FQ moment 
$O_{yz}$+$O_{zx}$+$O_{xy}$, which naturally brings about a trigonal lattice 
distortion through the quadrupole-strain interaction. Their calculation 
results give a reasonable magnitude of the spontaneous strain being comparable 
to the result obtained by the thermal expansion measurements reported by 
Akatsu {\it et al.}~\cite{rf:Akatsu2002} 
%
%
However, the behavior of $c_{44}$ cannot be explained quantitatively by the octupolar model. The magnitude of a jump in $c_{44}$ at $T_{c1}$ = 1.6 K predicted by their model (0.92$\times$10$^{11}$ erg/cm$^{3}$) is smaller than the observed softening (2.5$\times$10$^{11}$ erg/cm$^{3}$) in the phase IV. The reason of this discrepancy may come from the neglect of the fluctuation effect in their model as pointed out by Kubo et al. Ideally, the second derivative of the free energy such as elastic constants shows a discontinuity when the phase transition is of second-order. However discontinuities are not observed in the elastic constants $c_{44}$ of Ce$_{0.75}$La$_{0.25}$B$_{6}$ in the vicinity of the transition to the phase IV. We suppose that the absence of the discontinuity in the elastic constants is due to the quadrupolar fluctuation with $\Gamma_{5}$ symmetry. The existence of the quadrupolar fluctuation with $\Gamma_{5}$ symmetry is evident because the strong ultrasonic attenuation is observed in the echo signals of $c_{44}$. In addition, the inter-site octupolar interaction prevents the divergence of the quadrupolar susceptibility $\chi_{\Gamma_{5}}$, in other words, the elastic softening in $c_{44}$ stops at finite values not at zero, which is consistent with the experimental results of $c_{44}$. The octupolar model proposed by Kubo et al. can explain the elastic properties in the phase IV of Ce$_{0.75}$La$_{0.25}$B$_{6}$ as the mentioned-above qualitatively at least. This model is the most plausible model among the proposed ones of the ordering mechanism in the phase IV up to now, because the elastic constant, thermal expansion and the magnetic susceptibility can be explained naturally. Recent magnetic susceptibility measurements in uni-axial pressure by Morie {\it et.} {\it al.} also support the octupolar ordering model.~\cite{rf:Morie}
Furthermore, the breaking of the 
time reversal symmetry in the octupole ordering model may be consistent with the muon spin resonance and NMR experiment on the 
phase IV~\cite{rf:Takagiwa,rf:Magishi}. 
%
%
The octupole ordering of $T_{x}^{\beta}+T_{y}^{\beta}+T_{z}^{\beta}$ lifts the 
$\Gamma_{8}$ quartet into a singlet ground state, a first excited doublet 
state and a second excited singlet state. Hence the octupole ordered phase IV can 
be stable down to absolute zero. Actually in Ce$_{0.70}$La$_{0.30}$B$_{6}$ no phase transition except one at $T_{\rm c1}$ to the phase IV is observed  down to 50 mK~\cite{rf:Nemoto2002PhysicaB}.

    As shown in Fig. 8, the phase IV in the $H-T$ plane spreads as quite 
isotropic manner with respect to the magnetic field directions. The phase IV 
is easily destroyed by the weak magnetic field of 0.7 T. The AFM ordering 
temperature $T_{\rm c2}$ shifts to higher temperatures with increasing 
fields. The anisotropic field dependence of the AFM phase II is in contrast to 
the isotropic field dependence of the phase IV.
  The Kondo effect plays a role to screen the magnetic dipole moment in the AFM 
phases. Actually the tiny AFM component with the magnitude of about 0.25 $\mu_{\rm B}$ is observed in the 
phase III of Ce$_{0.75}$La$_{0.25}$B$_{6}$ by the neutron 
scattering~\cite{rf:Effantin1985JMMM}.  This is much smaller than the full 
moment of $\Gamma_{8}$ state ($\sim 1.57 \mu_{\rm B}$).  Furthermore, the 
recent measurements by Nakamura et al. showed the $T^2$-behavior in the 
temperature dependence of electrical resistivity in the phase IV of 
Ce$_{0.65}$La$_{0.35}$B$_{6}$~\cite {rf:NakamuraPRB2003}. This result 
suggests that the Fermi liquid state is realized by the dense Kondo effect 
in the phase IV of Ce$_{x}$La$_{1-x}$B$_{6}$.

\section{Summary}
 
We investigated the elastic properties of Ce$_{0.75}$La$_{0.25}$B$_6$ under 
magnetic fields by means of ultrasonic measurements. We obtained the 
magnetic phase diagrams of Ce$_{0.75}$La$_{0.25}$B$_6$ under fields along 
the three principal axes. The phase IV of Ce$_{0.75}$La$_{0.25}$B$_6$ 
is located only in the narrow temperature region. The phase IV spreads 
isotropically in $H-T$ plane being independent of the field directions. The 
huge softening in $c_{44}$ and the trigonal distortion of 
$\varepsilon_{yz}+\varepsilon_{zx}+\varepsilon_{xy}$ in the phase IV showed 
that the FQ moment of $O_{yz}$+$O_{zx}$+$O_{xy}$ is relevant in the phase 
IV. The antiferro octupole ordering model of 
$T_{x}^{\beta}+T_{y}^{\beta}+T_{z}^{\beta}$ proposed by Kubo and Kuramoto 
is the most plausible way to describe the elastic properties as well as the 
magnetic ones in the phase IV.

\newpage

{\bf Figure captions}

Fig.1. The elastic constants $c_{44}$ of Ce$_{0.75}$La$_{0.25}$B$_6$ as 
functions of temperature under various magnetic fields along the [001] 
axis. The data are cited from ref 11. Wave vector 
$\mib{k}$ and polarization one $\mib{u}$ of ultrasound are directed along 
the [001] axis and along the [100], respectively. Arrows are indications of 
phase transition points.

Fig. 2. The elastic constants $c_{44}$ of Ce$_{0.75}$La$_{0.25}$B$_6$ as 
functions of temperature under various magnetic fields along the  [110] 
axis. Wave vector $\mib{k}$ and polarization one $\mib{u}$ of ultrasound 
are directed along the [110] axis and along the [001], respectively. Arrows 
are indications of phase transition points.

Fig. 3. The elastic constants $c_{44}$ of Ce$_{0.75}$La$_{0.25}$B$_6$ as 
functions of temperature under various magnetic fields along the [111] 
axis. Wave vector $\mib{k}$ and polarization vector $\mib{u}$ of the 
ultrasound is directed along the [100] and [001] axes, respectively. Arrows 
are indication of the phase transition points.

Fig. 4. The elastic constant $(c_{11}-c_{12})/2$ of 
Ce$_{0.75}$La$_{0.25}$B$_6$as functions of temperatures under various 
magnetic fields along the [110] axis.  Wave vector $\mib{k}$ and 
polarization one $\mib{u}$ of ultrasound are  directed along the [110] axis 
and [1$\bar{1}$0] one, respectively. The origin of each data is shifted for 
clarity. Arrows are indication of the phase transition points.

Fig. 5. The elastic constant $c_{11}$ of Ce$_{0.75}$La$_{0.25}$B$_6$ as 
functions of temperatures under various magnetic fields along the [111] 
axis. Both the wave vector $\mib{k}$ and polarization one $\mib{u}$ of the 
ultrasound are directed along the [100] axis. Upward and downward arrows 
indicate $T_{\rm c1}$ and $T_{\rm c2}$, respectively. The origin of each 
data is shifted for clarity. 

Fig. 6. The elastic constant $(c_{11}-c_{12})/2$ of 
Ce$_{0.75}$La$_{0.25}$B$_6$as functions of magnetic field along the [110] 
axis at various temperatures.  Wave vector $\mib{k}$ and polarization one 
$\mib{u}$ of ultrasound are directed along the [110] and [1$\bar{1}$0] 
axes, respectively. The origin of each data is shifted for clarity. Arrows 
indicate phase transition points.

Fig. 7. The elastic constants $c_{44}$ of Ce$_{0.75}$La$_{0.25}$B$_6$ as 
functions of temperatures under various magnetic fields along the [110] 
axis. Wave vector $\mib{k}$ and polarization one $\mib{u}$ of ultrasound is 
directed along the [110] and along the [001] axes, respectively. Arrows 
indicate the transition points from the paramagnetic phase I to the AFQ 
phase II.

Fig. 8. The elastic constant $c_{11}$ of Ce$_{0.75}$La$_{0.25}$B$_6$ as 
functions of magnetic field  along the [111] axis under various magnetic 
fields. Both the wave vector $\mib{k}$ and polarization one $\mib{u}$ of 
the ultrasound are directed along the [100] axis. The origin of each data 
is shifted for clarity. Arrows are indication of phase transition points.

Fig. 9. Magnetic phase diagrams of Ce$_{0.75}$La$_{0.25}$B$_6$ in $H$-$T$ 
planes.
Magnetic fields are directed along the (a) [001], (b) [111], (c) [110] 
axes, respectively. The phase I, II and III are the paramagnetic phase, antiferro-quadrupolar phase and antiferro-magnetic phase in which antiferro-quadrupolar order coexists, respectively. The phase III' and III" are the sub-phase of the antiferro-magnetic phase III.

\begin{figure}[p]
\vspace{10pt}
\begin{center}
\includegraphics{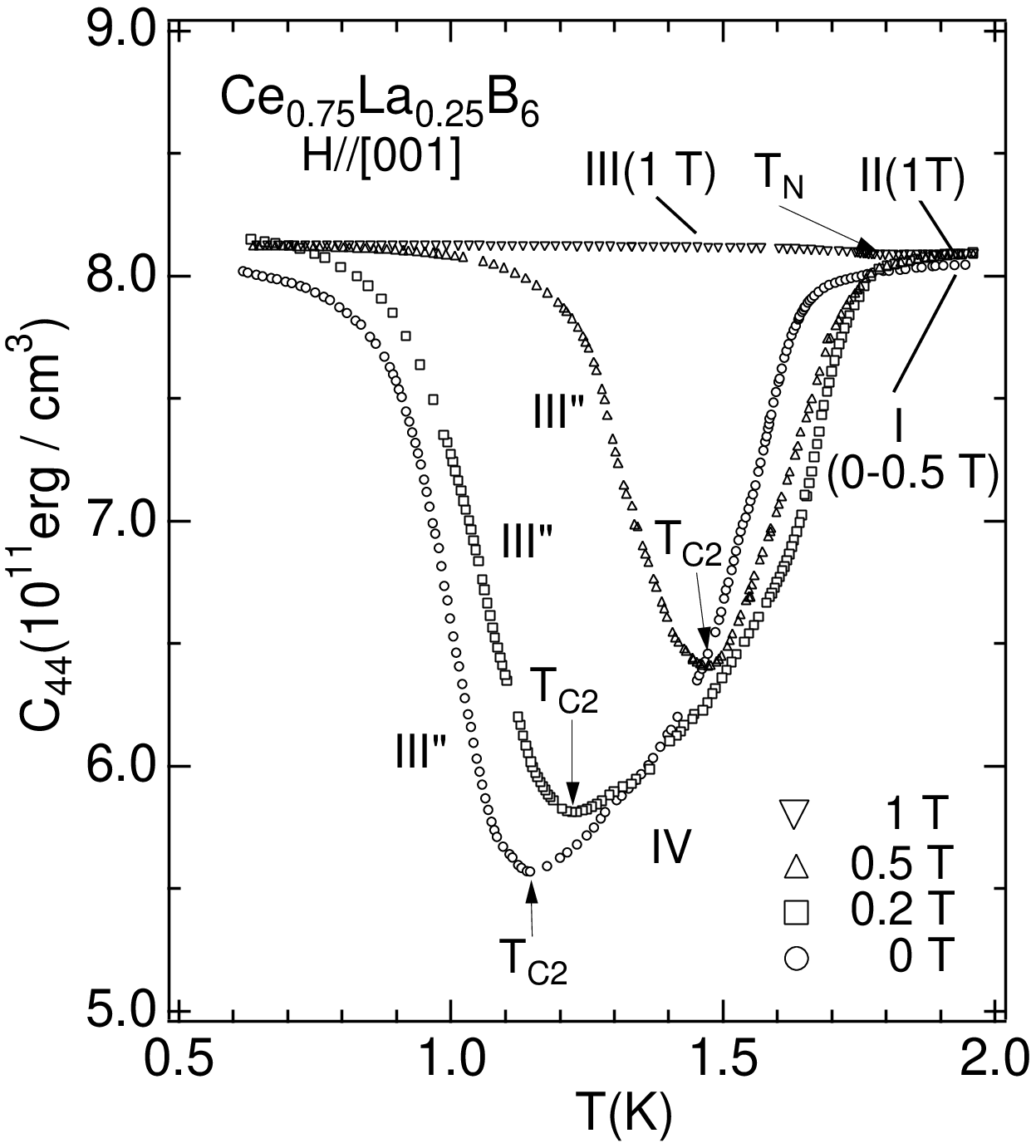}
\caption{}
\label{fig:1}
\end{center}
\end{figure}

\begin{figure}[p]
\vspace{10pt}
\begin{center}
\includegraphics{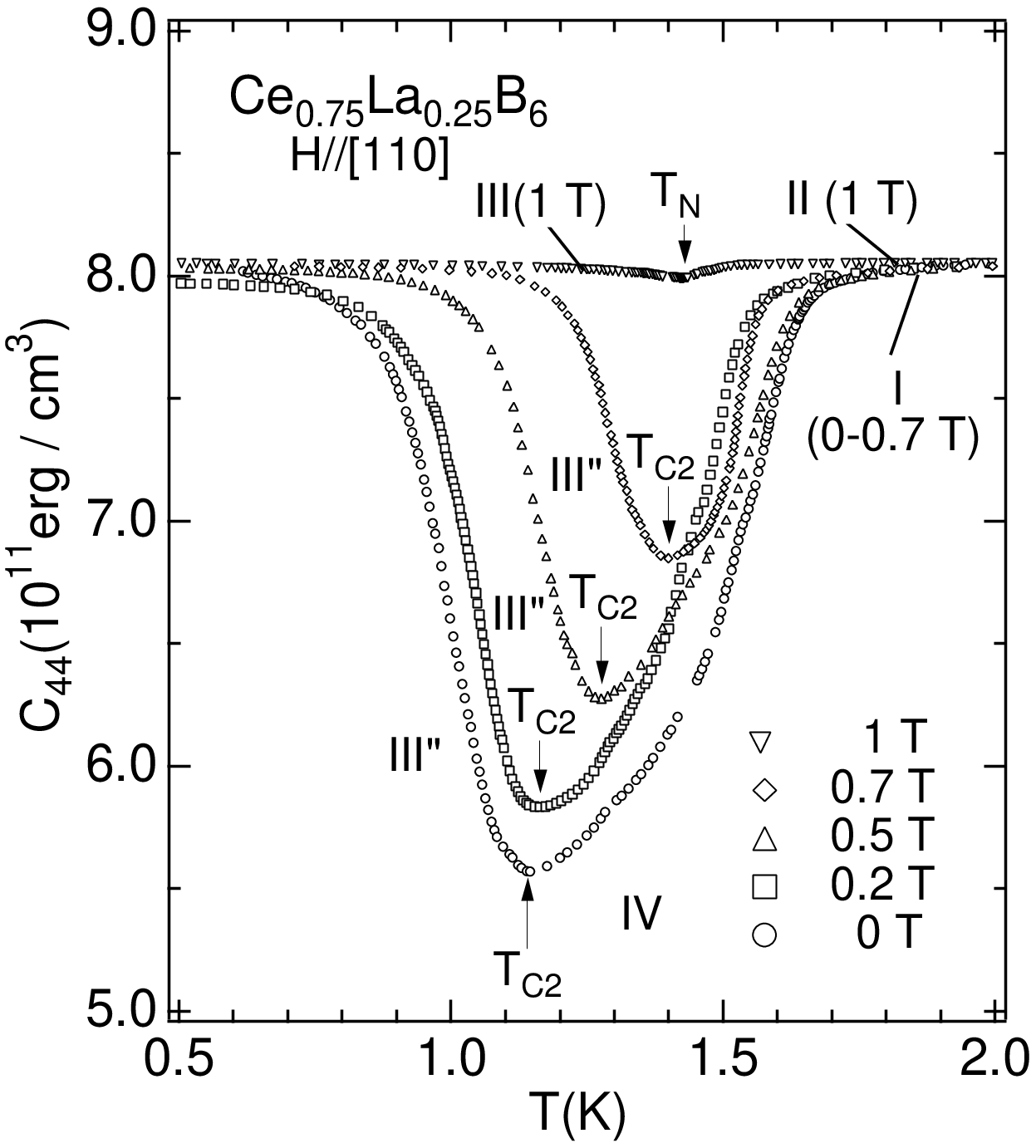}
\caption{}
\label{fig:2}
\end{center}
\end{figure}

\begin{figure}[p]
\vspace{10pt}
\begin{center}
\includegraphics{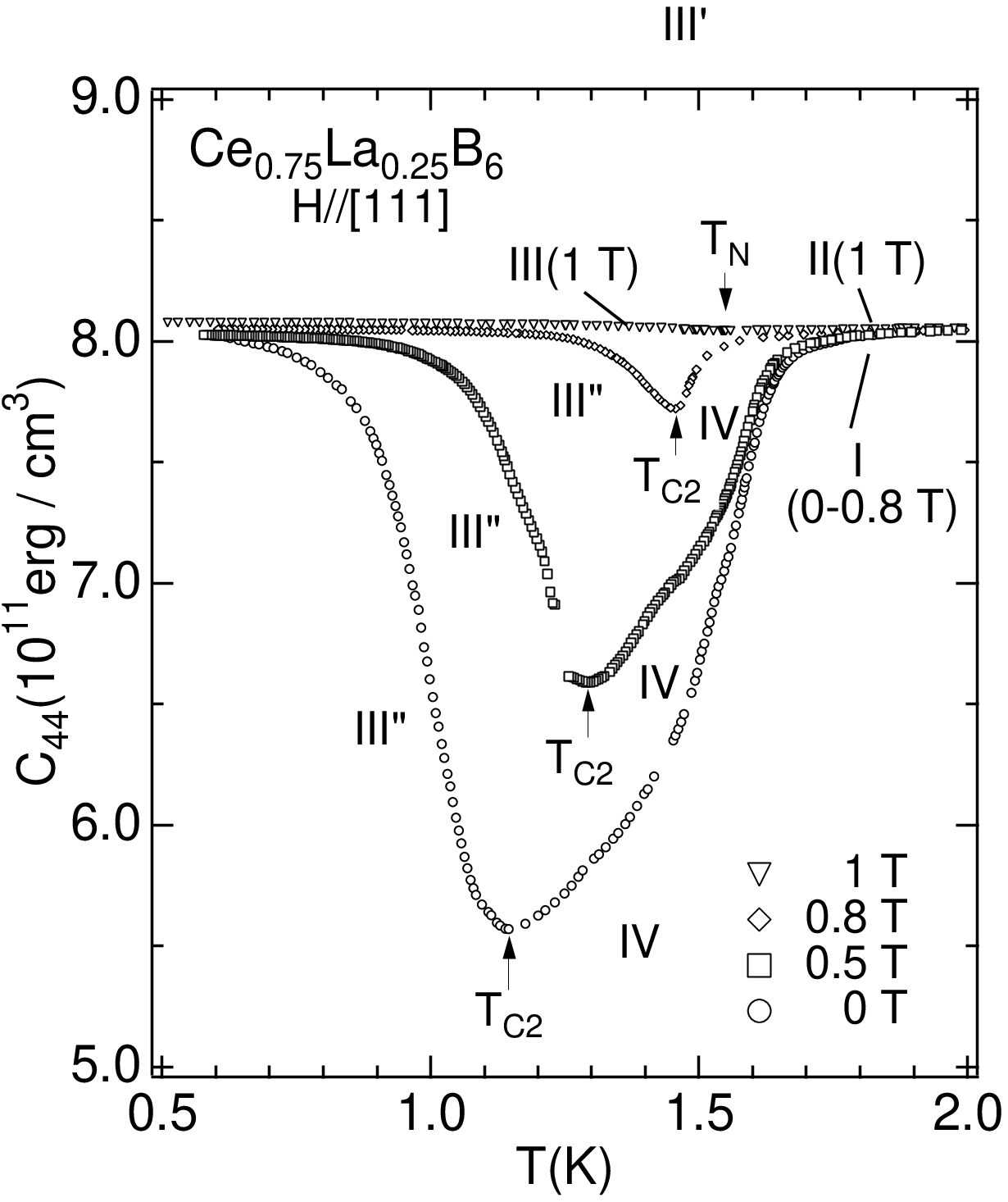}
\caption{}
\label{fig:3}
\end{center}
\end{figure}

\begin{figure}[p]
\vspace{10pt}
\begin{center}
\includegraphics{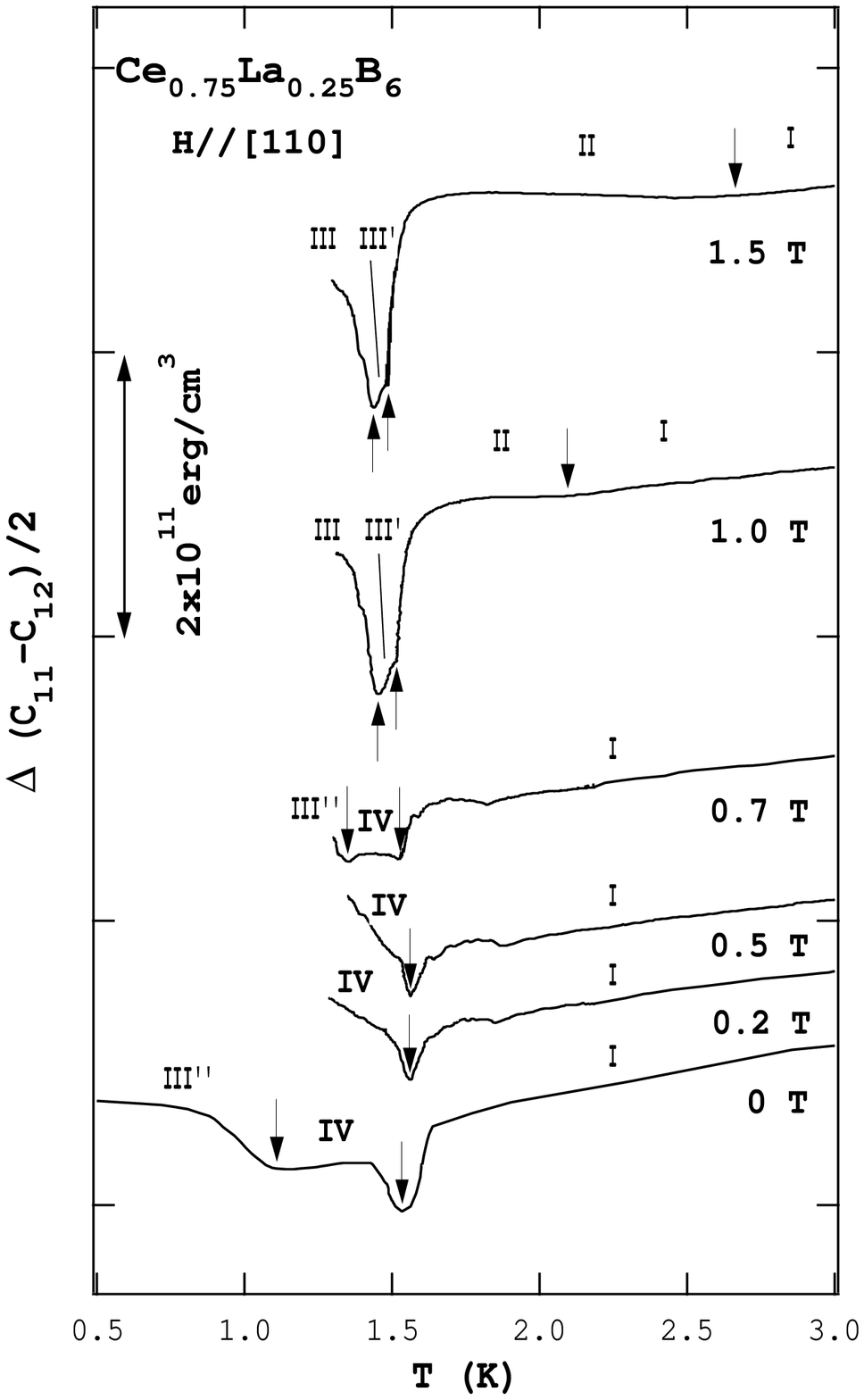}
\caption{}
\label{fig:4}
\end{center}
\end{figure}

\begin{figure}[p]
\vspace{10pt}
\begin{center}
\includegraphics{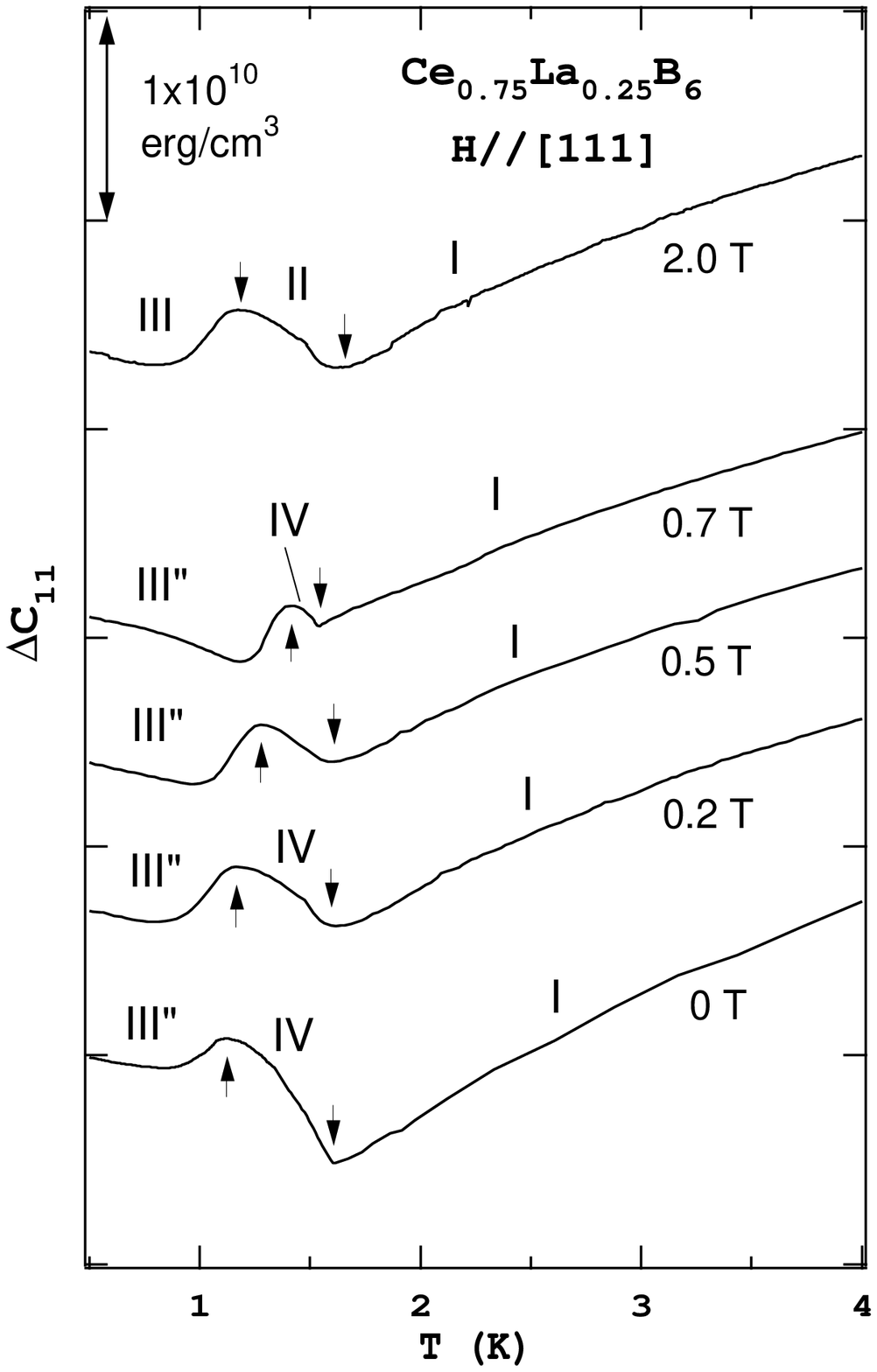}
\caption{}
\label{fig:5}
\end{center}
\end{figure}

\begin{figure}[p]
\vspace{10pt}
\begin{center}
\includegraphics{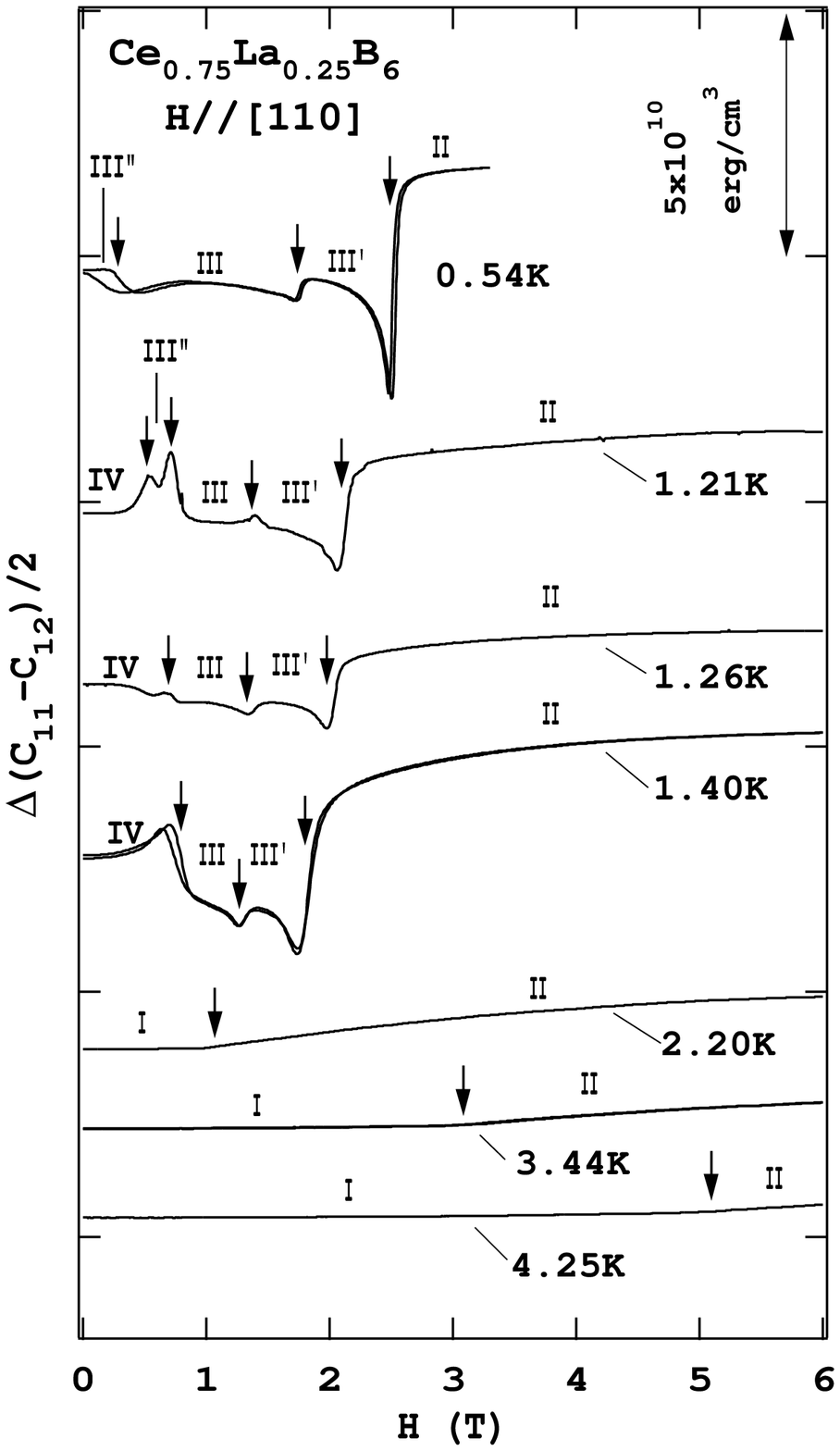}
\caption{}
\label{fig:6}
\end{center}
\end{figure}

\begin{figure}[p]
\vspace{10pt}
\begin{center}
\includegraphics{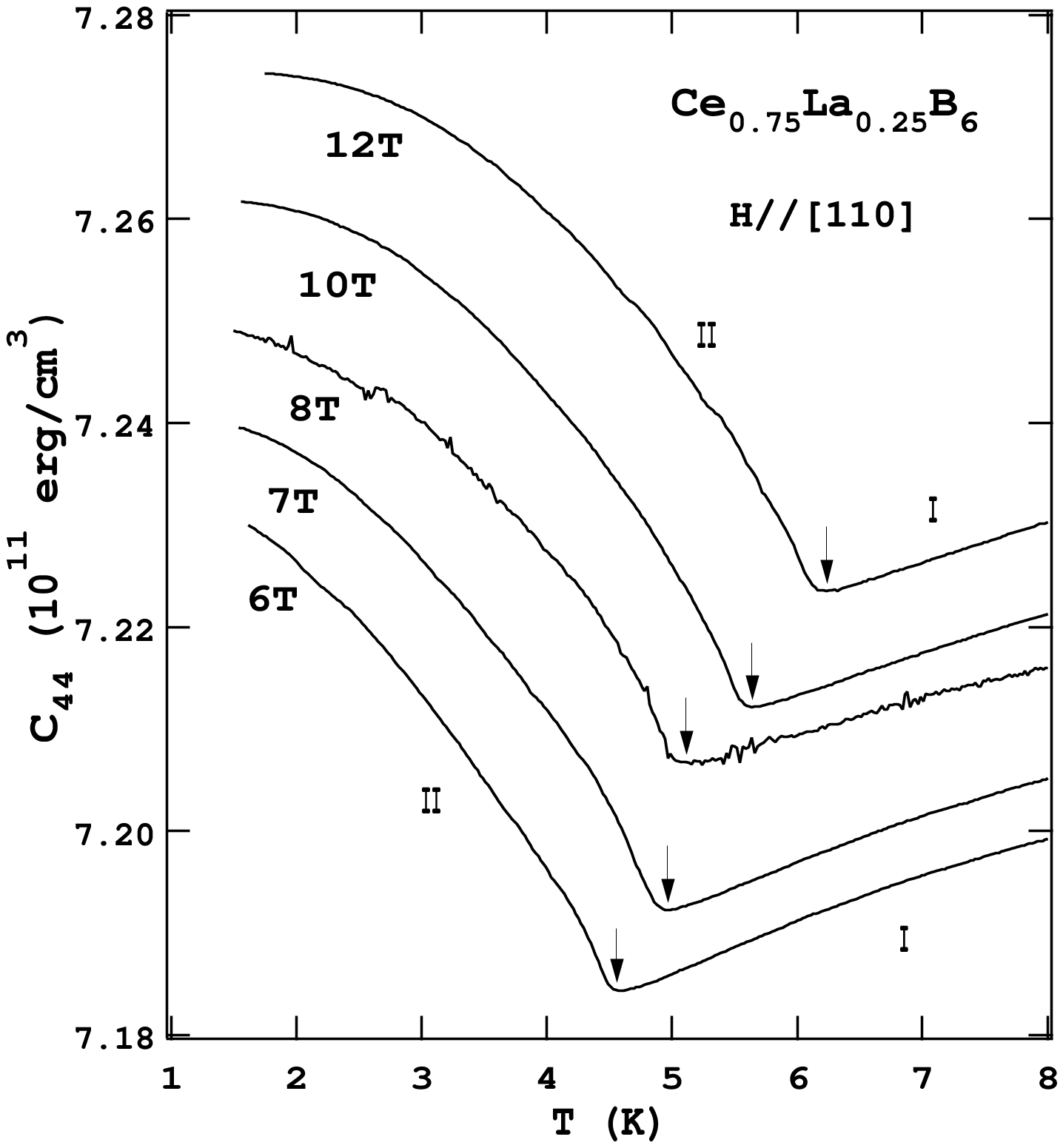}
\caption{}
\label{fig:7}
\end{center}
\end{figure}

\begin{figure}[p]
\vspace{10pt}
\begin{center}
\includegraphics{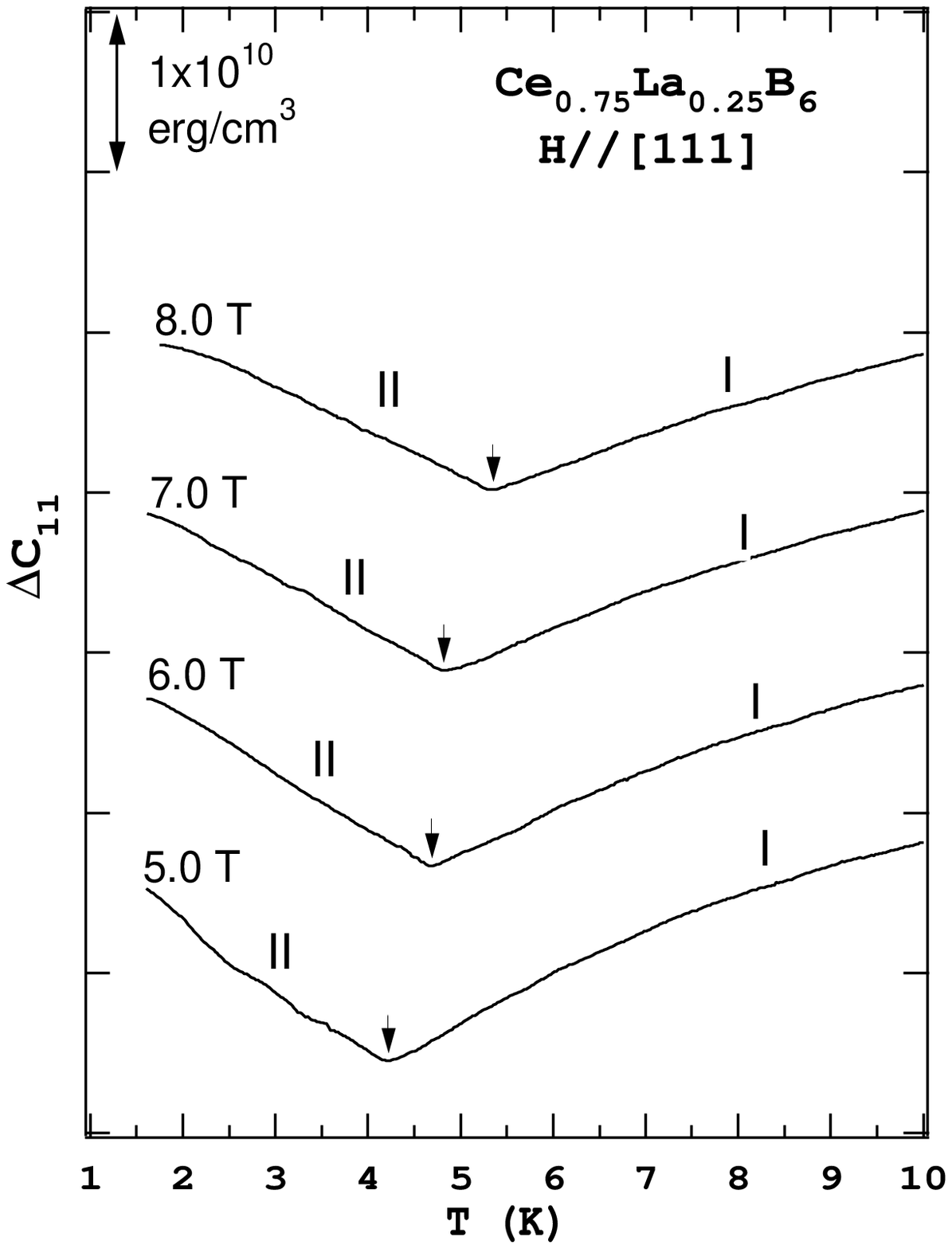}
\caption{}
\label{fig:8}
\end{center}
\end{figure}

\begin{figure}[p]
\vspace{10pt}
\begin{center}
\includegraphics[width=15cm]{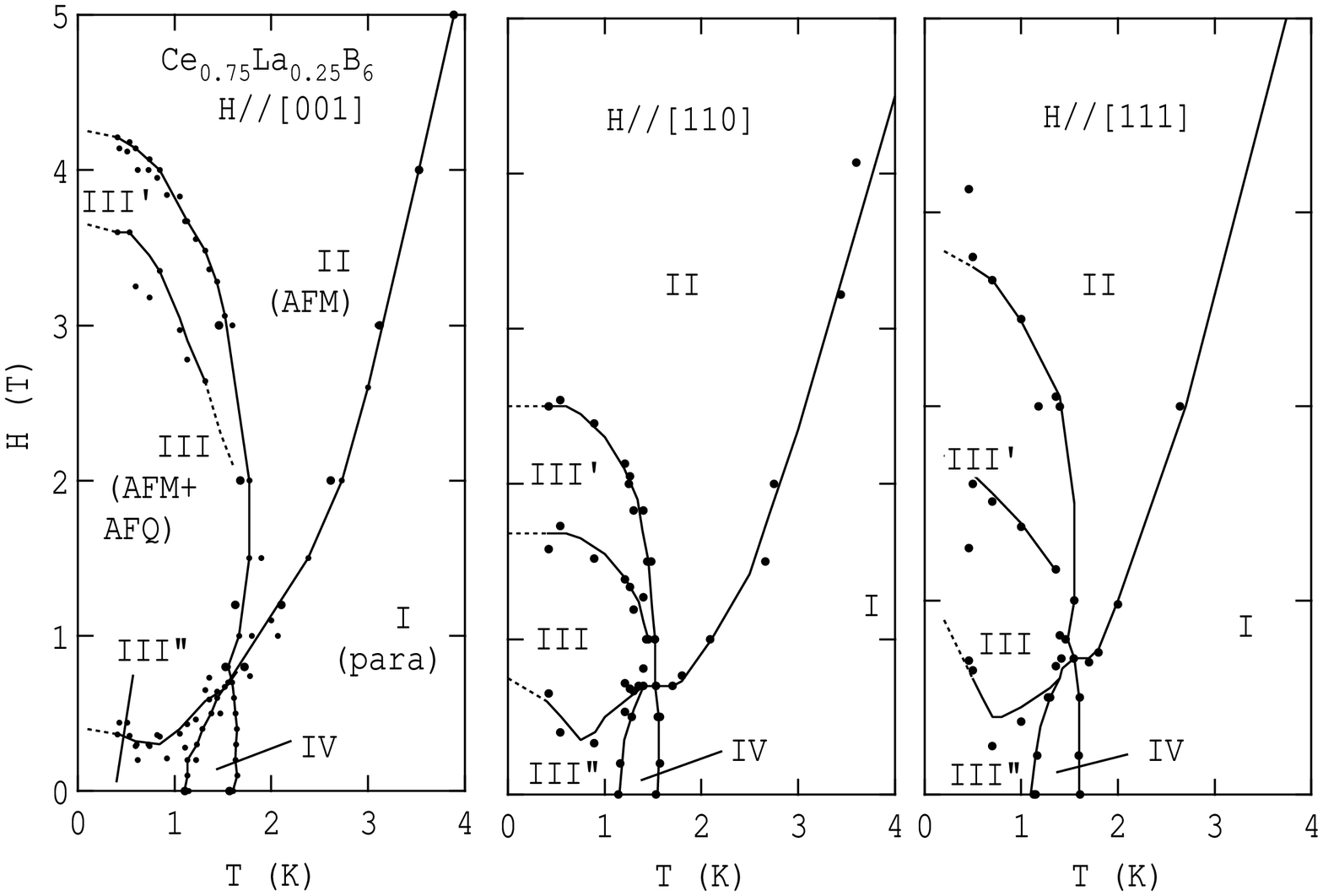}
\caption{}
\label{fig:9}
\end{center}
\end{figure}

\end {document}